\documentclass[aps,pre,twocolumn,a4paper,10pt,notitlepage,footinbib,superscriptaddress,showpacs,floatfix]{revtex4-1}%
\usepackage[english]{babel}
\usepackage[T1]{fontenc}
\usepackage{endnotes}
\usepackage{amssymb,amsmath,amsfonts}
\usepackage{textcomp}
\usepackage{graphicx,color}
\usepackage[dvipsnames]{xcolor}
\usepackage[utf8x]{inputenc}
\usepackage{float}

\usepackage{tabularx}

\renewcommand{\vec}[1]{\boldsymbol{\mathrm{#1}}}%

\begin{document}

\title{Plectoneme dynamics and statistics in braided polymers}


\author{Giada Forte}
\affiliation{Dipartimento di Fisica  e Astronomia and Sezione INFN, Università degli Studi di Padova, I-35131 Padova, Italy}
\author{Michele Caraglio}
\affiliation{Soft Matter and Biophysics section, KU Leuven, Celestijnenlaan 200D, 3001 Leuven, Belgium}
\author{Davide Marenduzzo}
\affiliation{SUPA, School of Physics and Astronomy, University of Edinburgh, Edinburgh EH9 3FD, United Kingdom}
\author{Enzo Orlandini}
\affiliation{Dipartimento di Fisica  e Astronomia and Sezione INFN, Università degli Studi di Padova, I-35131 Padova, Italy}

\date{\today}

\begin{abstract}
Braids composed of two interwoven polymer chains exhibit a "buckling" transition whose origin has been explained through the onset of plectonemic structures. Here we study, by a combination of simulation and analytics, the dynamics of plectoneme formation and their statistics in steady state. The introduction of an order parameter -- the plectonemic fraction -- allows us to map out the phase boundary between the straight braid phase and the plectonemic one. We then monitor the formation and the growth of plectonemes, observing events typical of phase separation kinetics for liquid-gas systems (fusion, fission, 1D Ostwald ripening), but also of DNA supercoiling dynamics (plectonemic hopping). Finally, we propose a stochastic field theory for the coupled dynamics of twist and local writhe which explains the phenomenology found with Brownian dynamics simulations as well as the power laws underlying the coarsening of plectonemes. 
\end{abstract}
\maketitle

\section{Introduction}

Polymer braids lie at a crossroad between topology, statistical mechanics, and biological physics. Within cells, two DNA double-helices inevitably intertwine to form a braided molecule, or catenane, when replication terminates~\cite{Mariezcurrena_et_al_G&D_2017}. Braids also arise in biology with interwoven collagen helices~\cite{Orgel_et_al_PNAS_2007}, or amyloid fibrils~\cite{Ionescu-Zanetti_et_al_PNAS_1999}. The biophysical properties of braided polymers can be quantitatively studied with single molecule experiments {\it in vitro}~\cite{Alfonsino_NAR_2014, Cejka_et_al_Mol_Cell_2012, Brahmachari_et_al_PRL_2017}. In nanotechnology, semiflexible polymers tethered to colloidal particles may be used to create biomimetic braids at near-molecular scale~\cite{Goodrich_Brenner_PNAS_2017}.

From a statistical mechanics viewpoint, these systems are of fundamental interest in view of the "buckling" transition between a \textit{straight braid phase}, where the braid centreline fluctuates around a straight line, and a \textit{plectonemic phase}, where the centreline writhes in 3D. The plectonemic phase arises for sufficiently large "catenation number" -- essentially the linking number between the two polymers in the braid. The thermodynamics of the transition is understood as a competition between torsional stresses stored in a twisted braid and bending energy cost associated with plectoneme formation, which can be formulated at the mean field theory level~\cite{Brahmachari_Marko_PRE_2017,Marko_PRE_1997,Neukirch_Marko_PRL_2011}. The underlying physics is similar to that of buckling in supercoiled DNA under tension, although in braids the absence of H-bonding between the intertwining molecules leads to important qualitative distinctions~\cite{Brahmachari_Marko_PRE_2017}.

A quantitative understanding of the transition between straight and plectonemic phase in braided polymers, going beyond mean field, is still elusive, although Monte-Carlo simulations have mapped out the phase diagram for DNA braids under tension, by monitoring the location of the singularity in the plots of braid extension versus catenation number~\cite{Charvin_et_al_2005}. The recent experiments in Ref.~\cite{Brahmachari_et_al_PRL_2017} also showed evidence of a multimodal distribution of braid extension (end-to-end distance), which arises because the population of braids in equilibrium has multiple plectonemes at large catenation. 

In sharp contrast to this body of work on the thermodynamics of the transition between the straight and plectonemic phase, the dynamics in the latter phase has received much less attention. Experimentally, the main study in this field has only addressed the diffusional dynamics of DNA {\it supercoils}~\cite{van_Loenhout_et_al_Science_2012}, whereas theory or simulation studies on braid dynamics are to date altogether lacking. To fill this gap, here we present Brownian dynamics (BD) simulations of two braided semiflexible polymers at fixed values of the catenation number and under a stretching force. 

After introducing the model of braided chains under torsion and the relevant observables (see Section II),  in section III we first map the  boundary between the straight and the plectonemic phase and then we focus on the  dynamics of the braid following a quench of variable depth into the latter. Our key finding is that the kinetics of plectonemic growth resembles phase separation in liquid-gas systems. Thus, first twist is converted into writhe to nucleate formation of a plectoneme. Later on, plectonemes coarsen via a combination of fusion and 1D Ostwald ripening. In steady state, we also observe plectonemic ``hopping'', whereby a plectoneme seemingly unravels at one place whilst reforming at another, as in the experiments of~\cite{van_Loenhout_et_al_Science_2012}. Finally in Section IV we propose a stochastic field theory for twist and writhe dynamics which explains the fundamental origin of the phase separation phenomenology we observe as due to the underlying topological conservation of the braid catenation number. Section V is devoted to discussion and conclusions.



\begin{figure}[!h]
\centering
\includegraphics[width=0.56\textwidth]{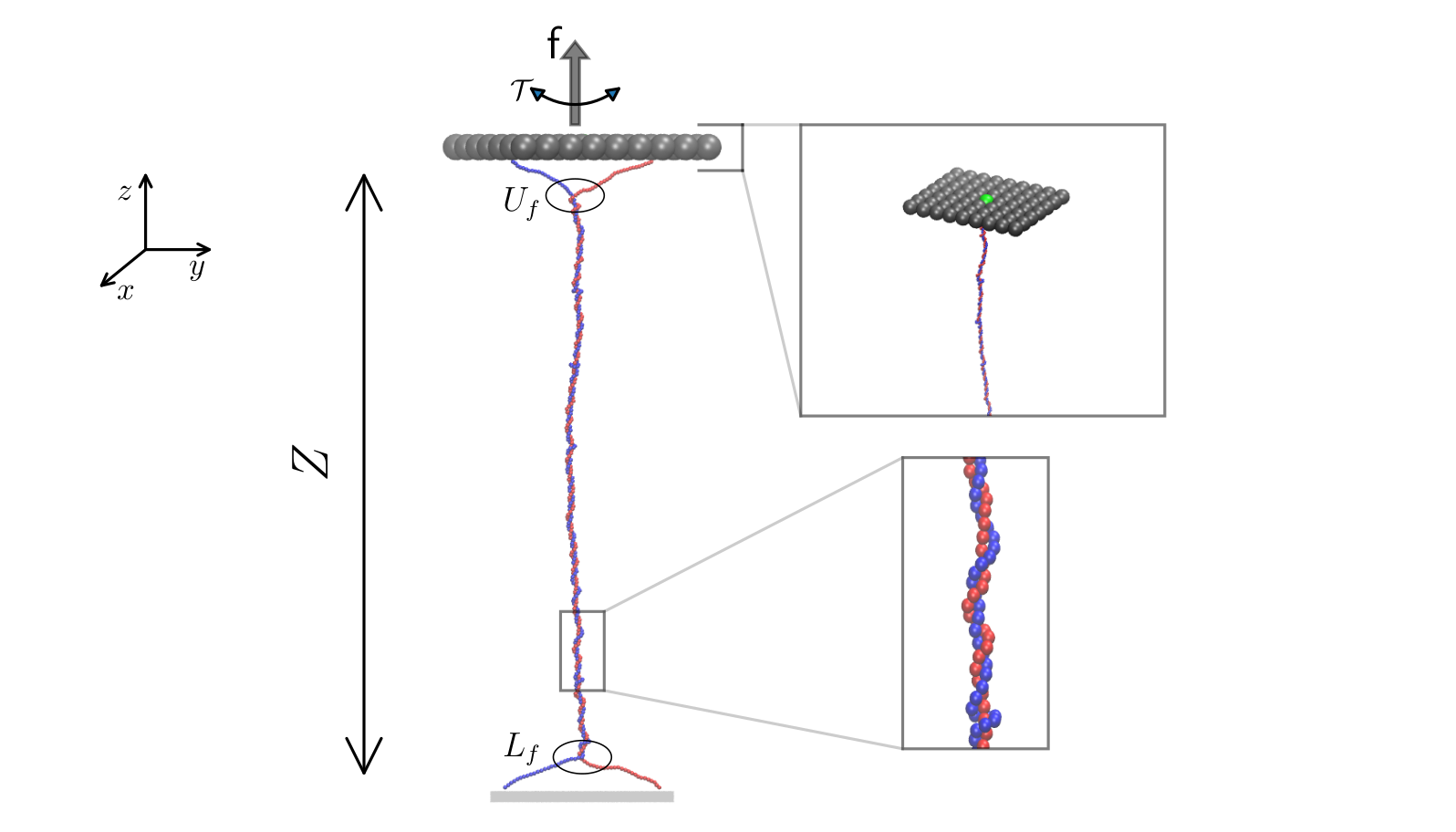}
\caption{\textbf{Setup of the molecular dynamic simulations}. The braid is made by two semiflexible chains. Each chain is composed by  $N=250$ beads with diameter $\sigma$ and mass $m$, and has a persistence length $l_p = 20.6 \, \sigma$. The two chains are anchored to a static impenetrable wall at the bottom and, at the top, to a rigid body composed of $63$ particles of radius $ 3.5 \, \sigma$. The rigid body is subjected to a force $f$ and a torque $\tau$ along the $z$-axis. We will refer to the braid extension $Z$ also as the \textit{end-to-end} distance. Finally $U_f$ and $L_f$ denote  the ends of the upper and lower fork respectively -- i.e., the points where the two
chains start to intertwine. \label{fig:S1}}
\end{figure}

\section{Model and simulation details}

Braids are modelled by a pair of semiflexible intertwined chains, each of which is made up by $N$ beads of diameter $\sigma$ and has persistence length $20.6\sigma$ (appropriate, e.g., for double-stranded DNA when $\sigma=2.5$nm) (see Fig.~\ref{fig:S1}). 
The total energy $U$ of each chain comprises three terms: (i) a finitely extensible nonlinear elastic (FENE) spring potential which accounts for the chain connectivity, (ii) a truncated and shifted Lennard-Jones potential which accounts for excluded volume interactions~\cite{Kremer_Grest_JCP_1990} and (iii) a bending potential which provides the chain with an intrinsic bending stiffness:
\begin{eqnarray}
&&U_{\mbox{\scriptsize{FENE}}}  = - \dfrac{K_{\mbox{\scriptsize{f}}}}{2} \sum_{i}^{N-1}  \left( \dfrac{R_0}{\sigma} \right)^2 \, \ln \left[1- \left( \frac{r_{i, i+1}}{R_0} \right)^2 \right] \; , \\
&&\frac{U_{\mbox{\scriptsize{LJ}}}}{4\epsilon}  = \sum_{i,j>i}^{N}  \left[ \left(\frac{\sigma}{r_{i,j}} \right)^{12} - \left(\frac{\sigma}{r_{i,j}} \right)^6 + \dfrac{1}{4}\right] \theta(2^{\frac{1}{6}} \sigma - r_{i,j}) \; , \\
&&U_{\mbox{\scriptsize{BEND}}} = \sum_{i=2}^{N-1} K_{\mbox{\scriptsize{b}}} \left[1 + \cos \, \phi_i \right] \; , 
\end{eqnarray}
where $r_{i,j} = \left| \vec{r}_{i,j} \right| = \left| \vec{r}_j-\vec{r}_i \right|$  is the distance between the $i$-th and the $j$-th bead, $R_0 = 1.5 \, \sigma$ is the maximum bond length, $\epsilon = k_B T$ is the thermal energy of the system, $ K_{\mbox{\scriptsize{f}}} = 30 \frac{\epsilon}{\sigma^2}$ is the bond strength, $\theta(x)$ is the Heaviside function and $\phi_i$ is the angle between the three consecutive beads $i-1$, $i$ and $i+1$. The value of $K_{\mbox{\scriptsize{bend}}}$ is related, in units of $\sigma$, to the persistence length of the chain  $l_p \simeq K_{b} / \epsilon$. By using $ K_{b} \approx 20 \epsilon$ we set $l_p =20.6 \, \sigma$. 
Note that beads belonging to different polymers also interact via excluded volume interactions:
\begin{equation}
\frac{U'_{\mbox{\scriptsize{LJ}}}}{4\epsilon} = \sum_{i,j}^{N} \left[ \left(\frac{\sigma}{r_{i,j}} \right)^{12} - \left(\frac{\sigma}{r_{i,j}} \right)^6 +\frac{1}{4} \right] \theta(2^{\frac{1}{6}} \sigma - r_{i,j}) \; ,
\end{equation}
where now $i$ and $j$ are two indices which are running on the first and second chain respectively.
To mimic a typical experimental magnetic tweezers set up,  each chain is anchored to an impenetrable wall at the bottom while the other pair of ends are fixed to the surface of a top plate that can rotate around the $z$-axis under the action of a constant torque $\tau$. 
Interactions between the bottom wall and the chains are accounted by the following harmonic potential:
\begin{equation}
U_{\mbox{\scriptsize{bot}}} = K_{\mbox{\scriptsize{w}}} (r-r_c)^2 \; ,
\end{equation}
where $r$ is the distance between the bead and the region surface, $r_c = \sigma$ is a cutoff distance below which there is no  interaction and $K_{\mbox{\scriptsize{w}}} = 200 \frac{\epsilon}{\sigma}$ is the spring constant.
The top plate  is composed by beads (labelled w) of diameter  $7 \, \sigma$ which interact  with the the beads in the two chains  (labelled b) through a truncated and shifted Lennard-Jones potential:
\begin{equation}\label{wallbead}
\resizebox{.9\hsize}{!}{$\frac{U_{\mbox{\scriptsize{top}}}}{4\epsilon} = \sum_{\lbrace b,w \rbrace} \left[ \left(\frac{\sigma}{d_{b,w}- \Delta} \right)^{12} - \left(\frac{\sigma}{d_{b,w} -\Delta} \right)^6 + \frac{1}{4}\right] \theta(2^{\frac{1}{6}} (\sigma + \Delta) - d_{b,w}) \; ,$}.
\end{equation}
In Eq.~\ref{wallbead} the sum runs over all pairs $(b,w)$ with reciprocal distance  $d_{b,w}$ and  $\Delta = R-\frac{\sigma}{2}$ with $R$ being the radius of the wall bead.
Note that in  all simulations we set the distance $d$ between the two end beads anchored at the top plate to be $d/L_0 \simeq 0.17$ where $L_0=N\sigma$ is the contour length of each chain~\cite{Nota1}.

The dynamics of the system is described by the Langevin equation:
\begin{equation}\label{bd}
m \ddot{\vec{r}}_i = - \gamma \dot{\vec{r}}_i + \vec{\xi}_i + \vec{\nabla}_{\vec{r}_i} U(\vec{r}_i) \; ,
\end{equation}
where $\vec{r}_i$ is the position of the \textit{i}th bead, $\gamma$ is the friction coefficient and $\vec{\xi}_i$ is the usual stochastic Brownian noise defined by 
\begin{eqnarray}
\langle \xi_{i,\alpha}(t) \rangle&=&0,\nonumber\\
\langle \xi_{i,\alpha}(t) \cdot  \xi_{j,\beta}(t') \rangle&=& 2\gamma k_B T \delta_{i,j} \delta_{\alpha,\beta} \delta(t-t') \: . \label{noise}
\end{eqnarray}
In Eq.~\ref{noise}, $t$ and $t'$ are times, $k_B$ is the Boltzmann constant, $T$ denotes temperature, $\delta_{i,j}$ is a Kronecker delta, and $\delta(t-t')$ a Dirac delta.
As characteristic time unit we consider the Brownian time $\tau_{LJ}= \frac{\sigma^2}{2D}$ (the time needed by a bead to diffuse across a length equal to its diameter): by using the Einstein relation $D= k_B T / \gamma$ and the Stoke's law $\gamma = 3 \pi \sigma \eta$ we obtain $\tau_{LJ} = \frac{3\pi}{2} \frac{\sigma^3 \eta}{k_B T}$ where $\eta$ is the viscosity of the fluid inside the flow cell. The time evolution of the system is obtained by integrating numerically Eq.~\ref{bd} with the \small{LAMMPS} software~\cite{plim95} with an integration time step in the range $[0.0025,0.01]\, \tau_{LJ}$. 

To simulate the system at equilibrium in the fixed catenation number (or inter-chain linking number) $Ca$ and force $f$ ensemble, initial conditions are prepared by interwining the two chains with the desired value of $Ca$. The latter is monitored by first closing the two chains and then computing their linking number. The system is then relaxed by BD simulations at  constant temperature $T$ and reduced force $\tilde{f} = f \frac{\sigma}{k_BT}$. Since chains cannot cross the top plate and the bottom wall, the initial catenation $Ca$ is preserved at all times. 



\begin{figure}[t]
\centering
\includegraphics[width=0.48 \textwidth]{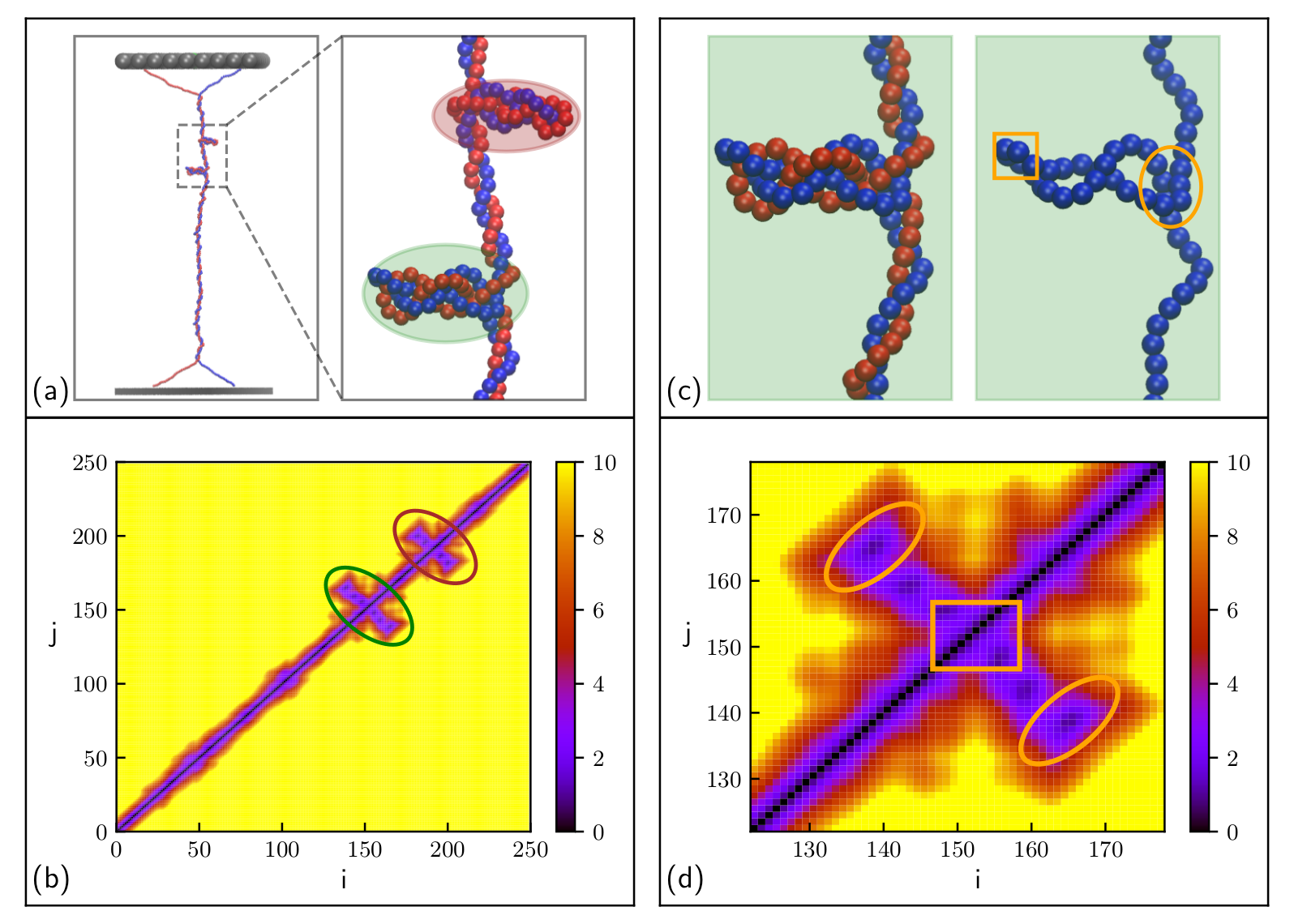}
\caption{\textbf{Detection of plectonemic domains in braids.}
\textbf{(a)} Snapshot of a braid in the plectonemic phase and a zoom showing its two plectonemes.
\textbf{(b)} Contact map referring to the blue chain of panel (a). The color map provides the distance between the \textit{i}th and the \textit{j}th bead in $\sigma$ units with a cutoff at $r_{cut}= 10 \, \sigma$.
The two plectonemes are described by the two darker regions moving away from the diagonal: the region within the brown (green) ellipse corresponds to the plectoneme highlighted by the brown (green) ellipse in panel (a).
\textbf{(c)} Zoom of the plectonemic region encircled by the green ellipse in panel (a). In the left snapshot one observes the whole braid, while in the right one only one chain is shown to better highlight  the bending of the chains in the plectoneme. 
\textbf{(d)} Zoom on the darker region within the green ellipse in panel (b) corresponding to the plectoneme represented in panel (c). The portions within the orange rectangle and ellipses correspond to the plectonemic areas of panel (c) enclosed in the rectangle and ellipse respectively. The rectangles include the apex of the plectoneme while the ellipse shows the location of its root.}\label{fig:S2}
\end{figure}

A crucial point in our study concerns the detection of plectonemes along the braided filaments (see Fig. \ref{fig:S2}a).
This is performed by first constructing a distance map for each of the two chains. This map is a contour plot of the distances between beads \textit{i} and \textit{j} of one chain (e.g. the blue one). [The distance map in Fig.~\ref{fig:S2}a has a cutoff at $r_{cut}= 10 \, \sigma$ ~\cite{Coronel_et_al_NAR_2018} as differences in distances of non-contacting monomers are irrelevant.] 
Plectonemic domains can be identified as darker regions extending perpendicularly from  the diagonal in the distance map (Fig.~\ref{fig:S2}b). The portion of these regions  closer to and farther from the diagonal represents respectively the apices and the roots of the plectonemes  (see Fig.~\ref{fig:S2}(c,d)).

More precisely, the algorithm to detect plectonemes through distance maps works as follows. First, we use a threshold $R_{\mbox{\scriptsize{max}}}$ to identify the set of  ``contacting beads'', for which $r_{i, j} \leq R_{\mbox{\scriptsize{max}}}$ (these are the darker regions in the maps in Fig.~\ref{fig:S2}b).  Then, we measure the positions of the plectoneme extremities (i.e. the points in the ellipses in Fig.~\ref{fig:S2}d) and use these to determine the plectoneme length. In our algorithm, plectonemes retained in the subsequent analysis were required to be larger than an additional threshold $\Delta_p$ (measured in units of $\sigma$) \cite{Nota1}.

Note that every distance map refers to a single chain: this means that we study plectonemes in each filament separately. Sometimes a chain wraps around the other locally forming a solenoidal structure whose beads satisfy the conditions determining a plectoneme: in this case our method of counting plectonemes can provide $N_b$ plectonemes for the blue chain and $N_r \neq N_b$ plectonemes for the red one. To avoid an overestimate of the  number of plectonemes, we assume that this number is given by $\min(N_b,N_r)$.

\vskip 1.0cm

\section{Results}

\begin{figure}[!]
\centering
\includegraphics[width=0.45\textwidth]{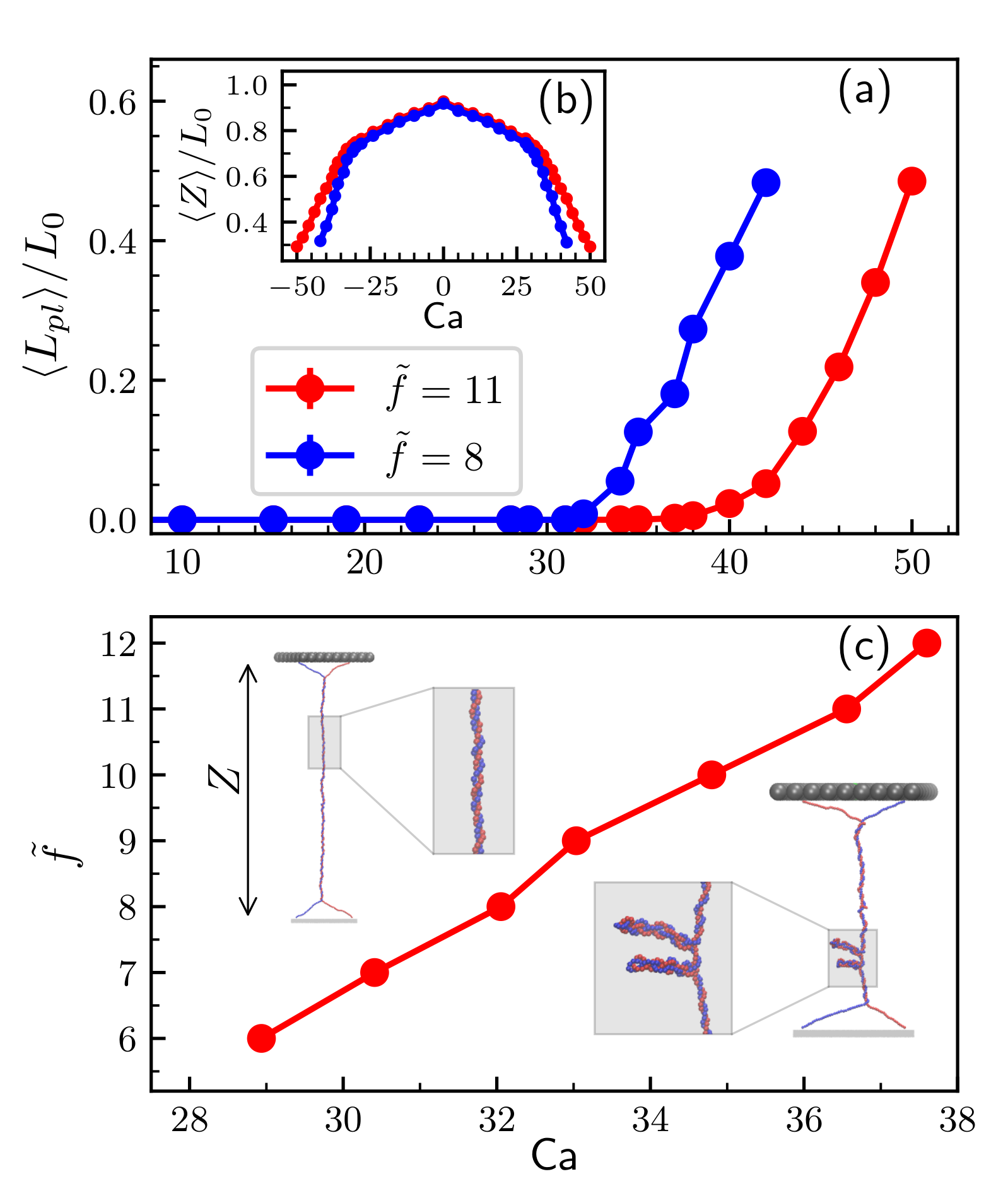}
\vspace{-0.3 cm}
\caption{\textbf{Thermodynamics of the transition between straight braid and plectonemic phase.} \textbf{(a)} Overall fraction of plectonemic length $\langle L_{pl}\rangle / L_0$ ($L_0 = N\sigma$ with $N=250$) as a function of $Ca$ for two values of $\tilde{f}$. $\langle L_{pl}\rangle / L_0$ becomes nonzero above a critical catenation number $Ca_c=Ca_c(\tilde{f})$, where plectonemes form. For each pair of $(Ca,\tilde{f})$ values, averages have been taken over $10$ trajectories.
\textbf{(b)} Mean normalised end-to-end distance $<Z>/L_0$  versus $Ca$ curves for the forces of panel (a).
\textbf{(c)} Numerical estimate of the equilibrium phase diagram in the $(Ca,\tilde{f})$ plane. The transition line (red circles) between the straight braid and the plectonemic phase has been determined as the set of point at which the order parameter $\langle L_{pl}\rangle / L_0$ deviates from zero. Note that the range of forces considered was chosen as this leads to better plectoneme detection. For braids made up by two double stranded DNA molecules of thickness $2.5$ nm, these values correspond to $\sim 10-20$ pN, larger than in normal experiments \cite{Brahmachari_et_al_PRL_2017}. } 
\vspace{-0.5 cm}
\label{fig:fig1}
\end{figure}

\subsection{Equilibrium phase diagram}

The results obtained for different $(Ca,\tilde{f})$ pairs are summarized in Fig.~\ref{fig:fig1}.
In thermodynamic equilibrium, and over the range of parameters we analyse, the braided polymers can be in one of two phases (Fig.~\ref{fig:fig1}c). Either the chains wind around each other without coiling in 3D (\emph{straight braid phase}, see snapshot in Fig. \ref{fig:fig1}, left), or they end up in a configuration where some of the twist is converted into writhe, leading to a shrinking of the braid along the $z$ direction and the formation of a plectoneme where the backbone of the braid writhes around itself (\emph{plectonemic phase}, see snapshots in Fig.~\ref{fig:fig1}c, right). 

To determine the boundary between the straight braid and plectonemic phase, the standard approach is to record the end-to-end distance of the chains along the $z$ direction as a function of $Ca$ -- the point at which the slope changes abruptly then corresponds to the transition point (Fig.~\ref{fig:fig1}b)~\cite{Charvin_et_al_2005,Brahmachari_et_al_PRL_2017}. Here, we instead look at the statistics of plectonemes by using the algorithm described in Section II and define an overall plectonemic length, $L_{\rm pl}$. The braid is in the plectonemic phase if the order parameter $\langle L_{\rm pl}\rangle/L_0$ -- the "plectonemic fraction" -- is $>0$. 
The plectonemic fraction as a function of $Ca$ for different values of $\tilde{f}$ is shown in Fig.~\ref{fig:fig1}a. 

\begin{figure}[!]
\includegraphics[width=0.45\textwidth]{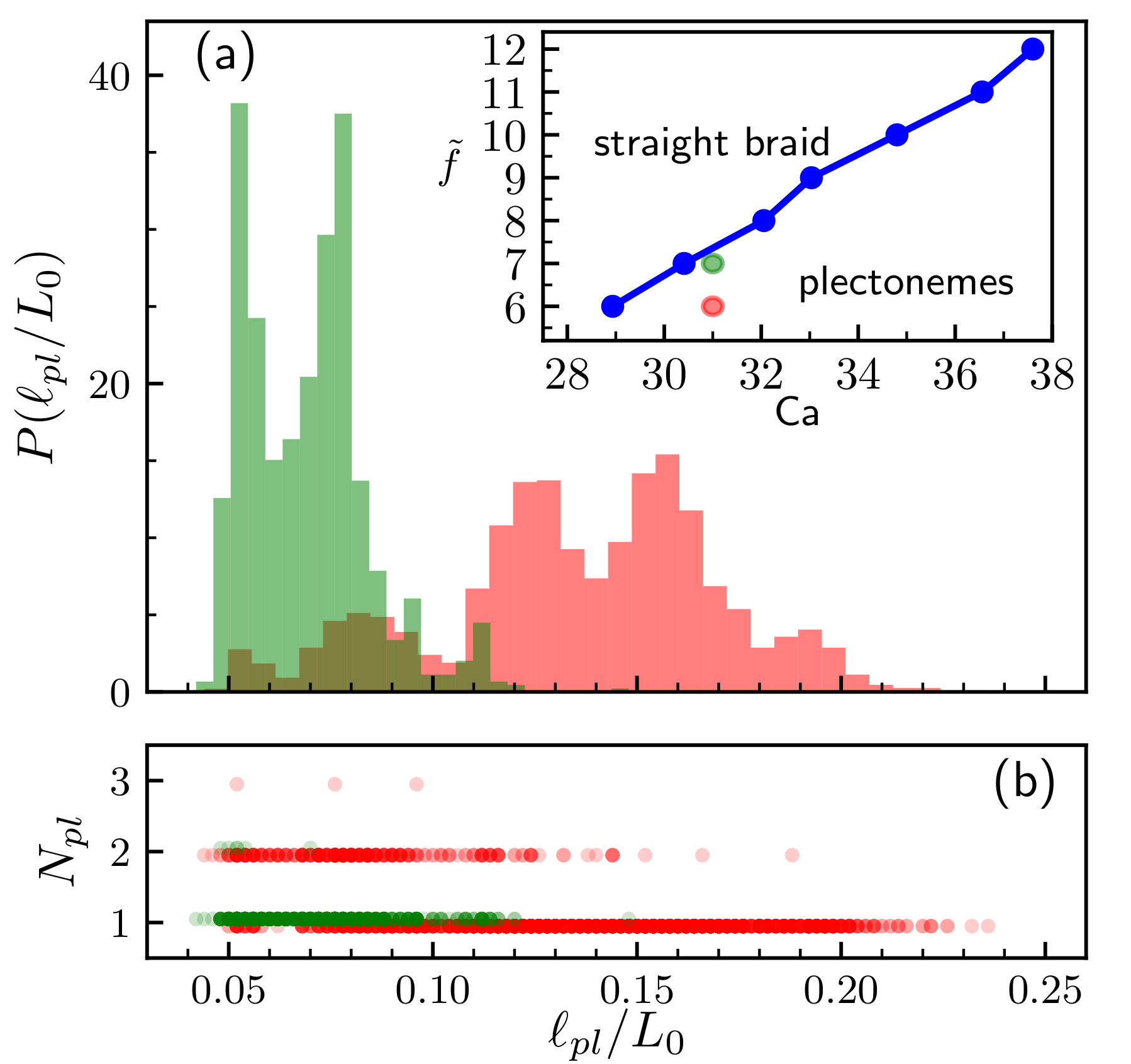}
\caption{\textbf{Plectoneme statistics.} 
\textbf{(a)} Equilibrium distribution of the relative length of a single plectonemes $P(\ell_{pl}/L_0)$ ($L_0 = 250 \sigma $), referring to different points of the phase diagram:  $(Ca = 31,\tilde{f}=6)$ (red) and $(Ca = 31,\tilde{f}=7)$ (green) (see inset). \textbf{(b)} Distribution of $\ell_{pl}/L_0$  as a function of the number of plectonemes $N_{pl}$. The statistics is based on  200 trajectories.
}
\label{fig:fig2}
\end{figure}

To gain more insight on the nature of the transition we look at the statistics of both the length $\ell_{\rm pl}$ of single plectonemes and their number $N_{\rm pl}$ in steady state. 
We find that the distribution of plectoneme length is multimodal (Fig.~\ref{fig:fig2}a) and that configurations with multiple plectonemes are possible (Fig.~\ref{fig:fig2}b). Close to the critical point, plectonemes are short and the length distribution is relatively narrow. Deeper in the plectonemic phase, the distribution widely broadens; additionally plectonemes are longer and more numerous. Multimodality in the length distribution here is much more enhanced and is due to the presence in the population of different classes of configurations, mainly those with a single, two or three plectonemes (Fig.~\ref{fig:fig2}b). Interestingly, multimodality in the plectonemic phase was recently reported in a combined experimental and theoretical study, but with respect to distribution of braid end-to-end distance~\cite{Brahmachari_et_al_PRL_2017}.

\subsection{Near-equilibrium dynamics of plectonemes}
We now discuss the near-equilibrium dynamics of the system after a quench from the straight braid phase into the plectonemic one by instantly reducing the pulling force. The kinetics depends on the value of $Ca$ and $\tilde{f}$. 


\begin{figure*}[!]
\centering
\includegraphics[width=0.95\textwidth]{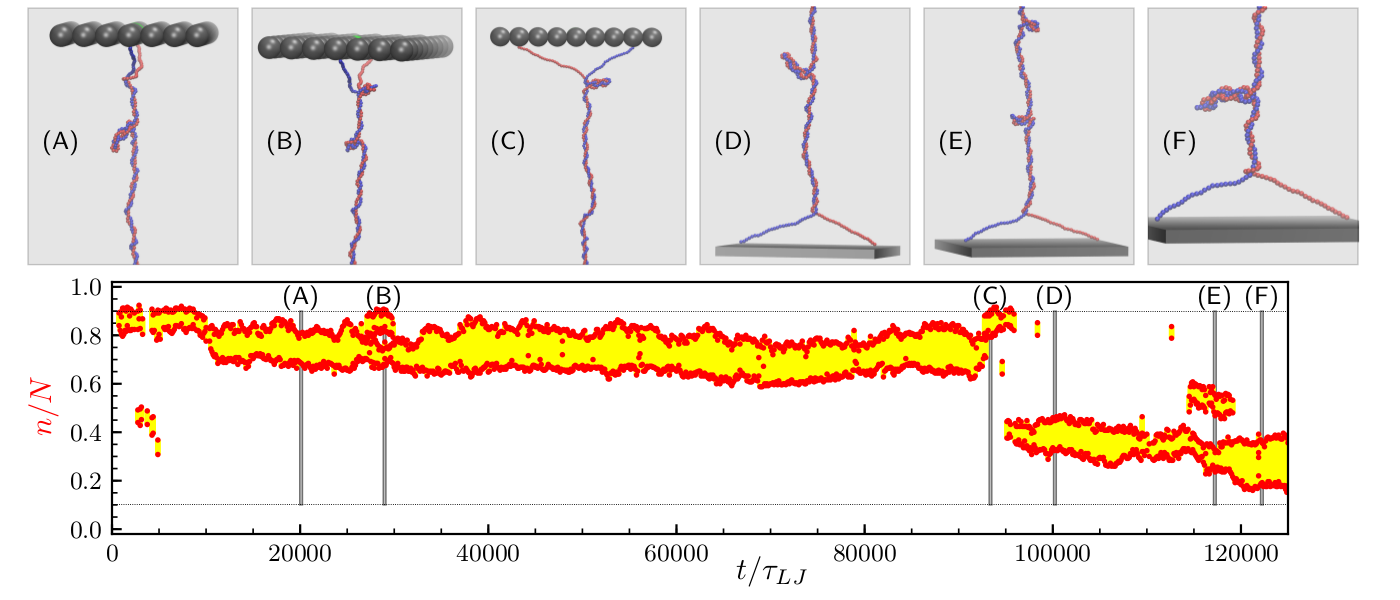}
\caption{\textbf{Plectoneme dynamics at equilibrium.} Kymograph of a pair of braided polymers, each of length $L_0=N\sigma=250\sigma$, with a fixed catenation number $Ca = 31$ and subject to a stretching force $\tilde{f} = 6$. On the $x$ axis one finds the time in Lennard-Jones units, while $n/N$ on the $y$ axis gives the relative position along the braid ($n/N=0$ and $1$ correspond to the bottom and top of the chains respectively). 
Plectonemes are visible as yellow regions bounded by red points and they always form within the two dashed lines representing the upper and the lower fork of the braid. The snapshots in panels (A-F) corresponds to specific events occurring in the plectoneme dynamics.
First we observe the growth of a single plectoneme which can diffuse, but also go through fission (between (A) and (B)) and fusion events (after (B)). A plectoneme can also hop from a position along the braid to another one far away ((C) to (D)). There is also evidence of 1D Ostwald ripening dynamics ((E) to (F)). }
\vspace{-0.6 cm}
\label{fig:fig3}
\end{figure*}

Following the quench into the plectonemic phase, the twist soon converts into writhe, with the latter initially localised close to the top plane (where the pulling force is applied). Then plectonemes can grow until they reach their equilibrium size. The kymograph in Fig.~\ref{fig:fig3} monitors the kinetics of plectonemes in the system when $(Ca, \tilde{f}) = (31,6)$. In our BD simulations, we observe a number of possible kinetic events. First, plectonemes diffuse slowly along the chain once they are formed. Second, there are  fission (Fig.~\ref{fig:fig3} A to B) and fusion (Fig.~\ref{fig:fig3} after B) events, where a plectoneme splits into two, or two close plectonemes merge into one. 
Third, plectonemes may "hop" from one place to another along the braid, as writhe unravels in one region and reappears in another one (see snapshots from Fig.~\ref{fig:fig3} C to D). Finally, we observe Ostwald ripening events (Fig.~\ref{fig:fig3} E to F) where a small plectoneme is absorbed by a larger one, without touching. 


Within the context of DNA supercoiling, plectonemic diffusion was observed in experiments~\cite{van_Loenhout_et_al_Science_2012} and simulations~\cite{Matek_Sci_Rep_2015}, and hopping only in experiments~\cite{van_Loenhout_et_al_Science_2012}. Inspection of our BD simulations (Suppl. Movie~1) suggests that hopping events are actually preceded by a fluctuation resulting in the moderate shrinkage of an initially large plectoneme. The extra ropelength gained allows nucleation of new small plectonemes, at random positions along the chain, which compete with the original plectoneme for braid length. If one of these outgrow the original plectoneme, the resulting kymograph records a hopping. As hopping requires nucleation, its frequency should strongly depend on noise -- this is the case in our simulations  and arguably also in experiments~\cite{van_Loenhout_et_al_Science_2012}.
It is then worth noting that fission and fusion events balance in steady state to yield a finite average size for plectoneme. This is similar to liquid-gas phase separation in 1D, where fluctuations inhibit coarsening beyond the correlation length of the system~\cite{Cates_PRL_2008}. 

\begin{figure*}[t]
\centering \includegraphics[width=0.95\textwidth]{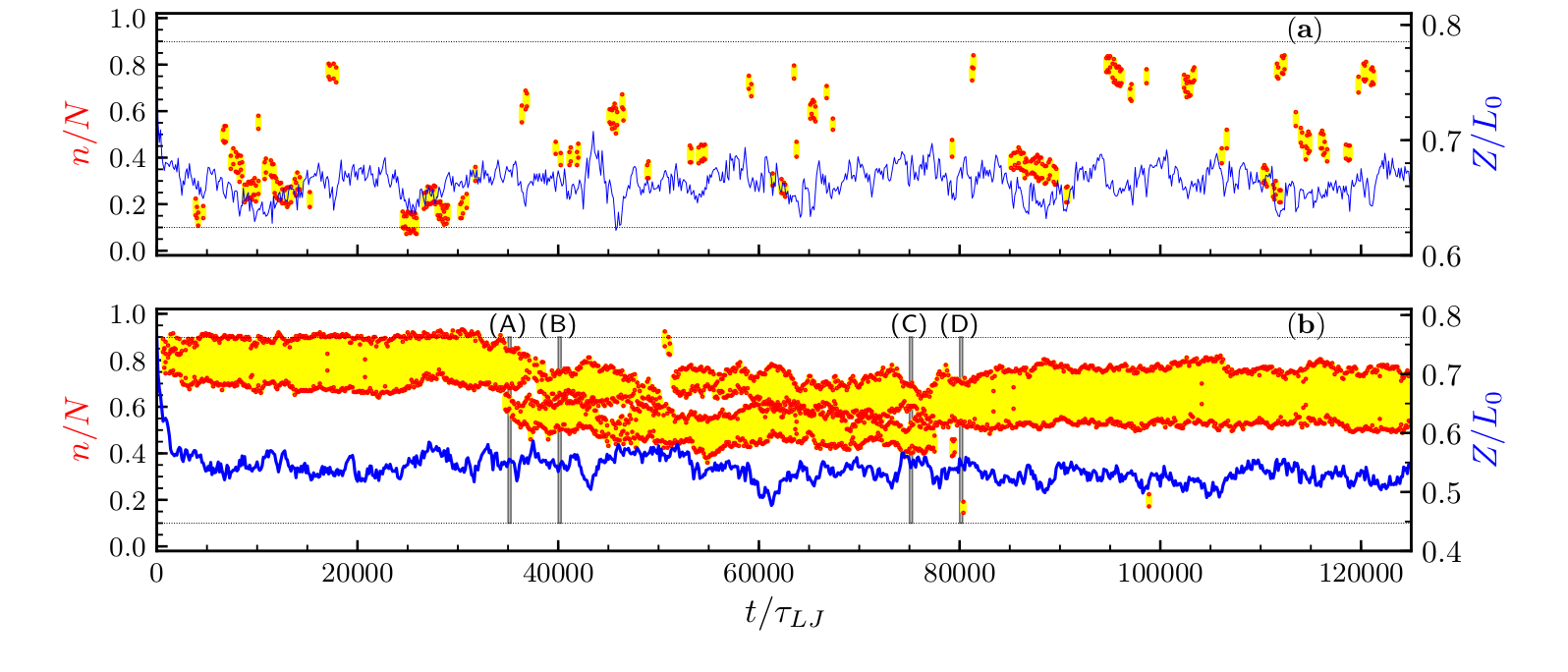}
\caption{
\textbf{Comparison between dynamics of plectonemes at equilibrium.}
\textbf{(a)} Kymograph of a simulation taken at the point $(Ca, \tilde{f}) = (32, 8)$. In the phase diagram this point is located close to the transition line. On the left \textit{y} axis the normalized plectonemic position $n/N$ is reported: yellow points correspond to beads within a plectoneme, while the red ones indicate its extremities. The blue curve describes the time evolution of the normalized end-to-end distance, $Z/L_0$ (see the right \textit{y} axis).  One can  notice the formation of a gas of small plectonemes  where each domain is enclosed within the lower and the upper fork of the braid (see black dotted lines).
\textbf{(b)} This kymograph refers to a trajectory simulated at $(Ca, \tilde{f}) = (32,6)$ i.e. 
well inside the plectonemic phase. In this case larger and more stable plectonemes are present. Notice events such as fission (A to B) and fusion (C to D) between plectonemes.
}
\label{fig:S4}
\end{figure*}

The plectoneme dynamics when the quench is deep in the plectonemic phase, compared to the one close to the transition, is very different. This is apparent in Fig.~\ref{fig:S4} where we contrast the trajectories for quenches to the point $(Ca, \tilde{f})= (32,8)$ (panel (a)) and for $(Ca, \tilde{f}) = (32,6)$ (panel (b)). One can notice that both trajectories follow a fast relaxation to equilibration dynamics  in which the observable $Z /L_0$ (blue line) drops abruptly (this is particularly apparent in panel (b)). This fast decay is due to the sudden reduction of the pulling force at $t=0$, leading to the transition from a straight braid to a braid where plectonemes nucleate.
\\
\noindent
Close to the plectonemic transition a  gas of plectonemes appears: numerous domains nucleate between the lower and the upper fork, but they do not merge as their lifetime is too short. The braid end-to-end distance is weakly influenced by the formation of domains and only occasionally its decrease is clearly visible due to  the presence of plectonemes.\\
Deep in the plectonemic phase instead the system prefers to form longer plectonemes. 
While the total length of plectonemes is practically constant their number varies. As in Fig.~\ref{fig:fig3}, we find events which are typical of liquid-gas phase separation, such as fission (Fig. ~\ref{fig:S4}b A to B) and fusion (Fig. ~\ref{fig:S4}b C to D).
Also in this case the extension of the braid is not strongly influenced by the nucleation of plectonemes and the size of fluctuations is similar to the one seen close to the transition.

\begin{figure}[!]
\includegraphics[width=0.43\textwidth]{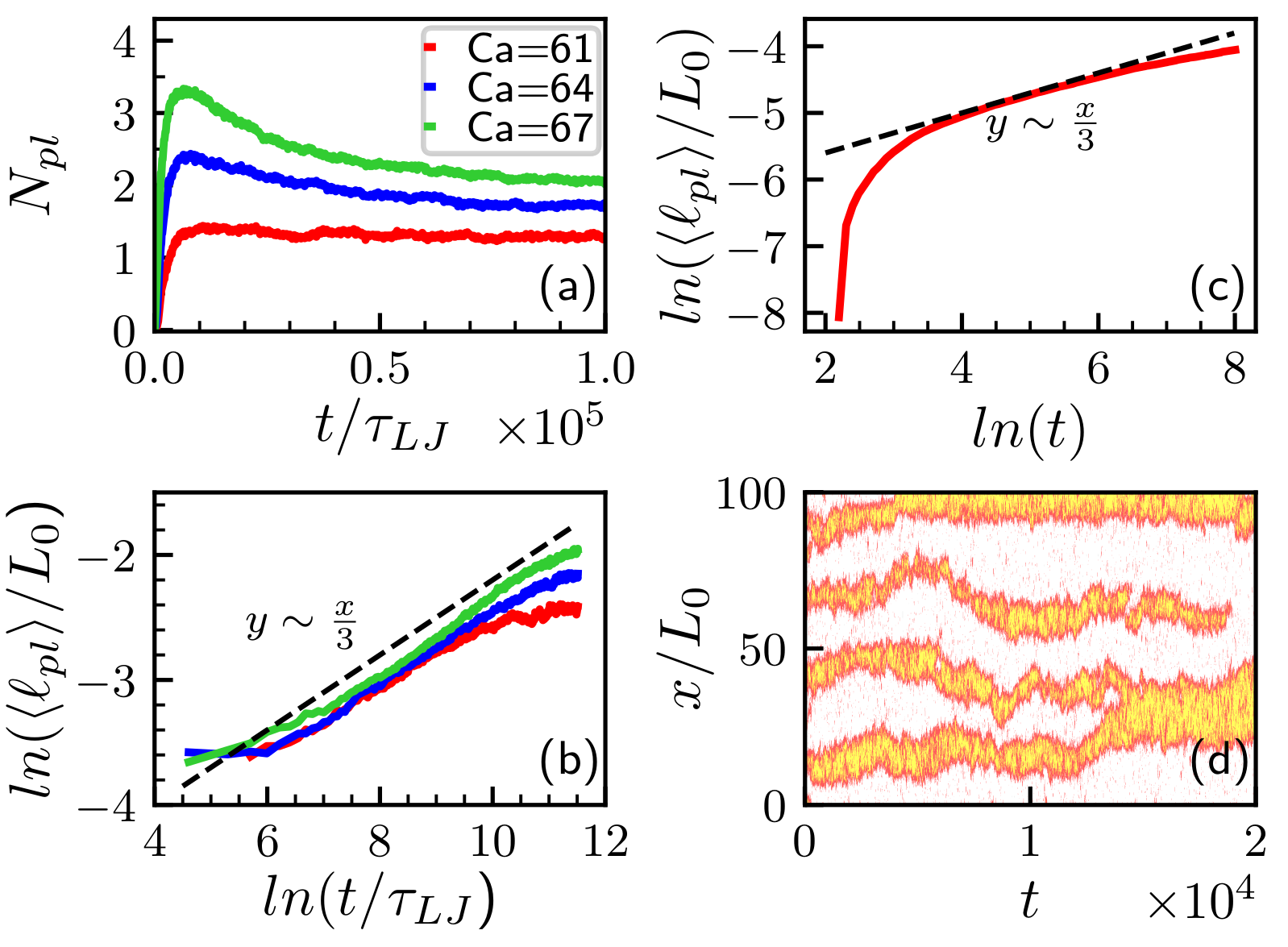}
\caption{\textbf{Plectoneme coarsening}. BD simulations (a,b) and  stochastic field theoretical model (c,d) allow to monitor the growth of plectonemes at intermediate times.    
\textbf{(a)} Variation of the number of plectonemes $N_{pl}$ over time following a quench from the straight braid phase with $\tilde{f}=9$ into the plectonemic phase with $\tilde{f}=6$. The value of $Ca$ is kept fixed as specified in the figure.
\textbf{(b)} Log-log plot of  $\langle \ell_{pl} \rangle /L_0$ ($L_0 = N \sigma=500\sigma$) versus $t/\tau_{LJ}$ for the quenches of panel (a). Curves of panel (a) and (b) are averaged over 250 trajectories.
\textbf{(c)} Time evolution of the mean length of a plectoneme obtained through the stochastic field theory (average over $50$ simulations). Parameters were $D_t=2$, $D_w=0.1$, $\alpha=\beta=0.1$, $\gamma=0.07$, $\kappa=0.01$, $\sigma_{\eta}=0.003$, $L_0=500$. This simulation and the associated power law behaviour are representative of the typical behaviour of the system. 
 \textbf{(d)} Kymograph corresponding to a simulation of the stochastic field theory with parameters as in (c), except for $\sigma_{\eta}=0.087$ and $L_0=100$. }

\label{fig:fig4}
\end{figure}

\subsection{Coarsening dynamics}
To characterise more quantitatively the dynamics of plectonemic coarsening following a quench, we analyse in Fig.~\ref{fig:fig4} how the average plectoneme size and number depends on time. The number is non-monotonic with a maximum that is reached at very early times and whose values increase with $Ca$ (see Fig.  \ref{fig:fig4}a). This behavior is due to the formation of numerous plectonemes just after the quench and to the following process of coarsening. That the number is larger than $1$ in steady state is consistent with arrested coarsening. The intermediate time behavior is well fitted by a power law $t^{\omega}$, with $\omega\simeq 1/3$ -- to a good approximation this is true irrespective of the value of $Ca$ and $\tilde{f}$ (Fig.~\ref{fig:fig4}b). 

\section{Stochastic field theory}
The phase separation kinetics between straight and plectonemic phases can rationalized via a simple stochastic field theory, based on two coupled reaction-diffusion equations for the evolution of local twist (${\rm Tw}$) and centreline writhe (${\rm Wr}$), inspired by our BD simulations. These equations read as follows (see appendix A  for more details),
\begin{eqnarray}\label{meanfieldtheory}
\frac{\partial {\rm Tw}}{\partial t} & = & D_t \frac{\partial^2 {\rm Tw}}{\partial x^2} - g({\rm Tw},{\rm Wr}) \\ \nonumber
\frac{\partial {\rm Wr}}{\partial t} & = & D_w \frac{\partial^2 {\rm Wr}}{\partial x^2} - \kappa \frac{\partial^4 {\rm Wr}}{\partial x^4}+g({\rm Tw},{\rm Wr}) \\ \nonumber
g({\rm Tw},{\rm Wr}) & = &  \alpha {\rm Tw}   - \beta {\rm Wr} + \gamma \theta\left(|{\rm Tw} + {\rm Wr}| - {\sigma}_0\right) {\rm Wr} \\ \nonumber &+& \sigma_{\eta} \eta(x,t),
\end{eqnarray}
where $D_t$ and $D_w$ are the diffusion coefficients of twist and (local) writhe respectively, while $\kappa>0$ is a surface tension-like parameter. The reaction terms $\pm g$ are equal and opposite in the two equations to ensure global conservation of the "local catenation" number, ${\sigma}={\rm Tw}+{\rm Wr}$, at all times. 
The parameters $\alpha$, $\beta$ and $\gamma$ describe interconversion between twist and writhe -- we choose $\alpha\ge \beta\ge \gamma>0$ to locally favour formation of a plectoneme for $\sigma>\sigma_0$; we also include a noise term, where $\sigma_{\eta}$ denotes noise strength and $\eta(x,t)$ is a Gaussian white noise, uncorrelated in space and time. Eq.~\ref{meanfieldtheory} (with no noise) is familiar in the literature on Turing pattern formation~\cite{Murray_Math_Bio_2001, Frey_Nat_Phys_2018}, where a sufficiently large difference in the diffusivity of the fast and slow fields (here twist and writhe respectively) leads to an instability of the uniform phase (Fig.~\ref{fig:S8} and Appendix). [In our context, the instability corresponds to phase separation and formation of one or more plectonemes in the braid.] For $D_t=D_w$, the linking number obeys a simple diffusion equation, hence is uniform in steady state.

Fig.~\ref{fig:fig4}c shows the evolution of the average plectoneme size in numerical solutions of our stochastic field theory. Due to the global conservation of $\sigma$ we expect Eq.~\ref{meanfieldtheory} to be in the same universality class of model B dynamics for liquid-gas phase separation (this is proved in the Appendix), for which we expect a growth exponent $\omega=1/3$ \cite{Bray_2002} -- as we find numerically in both BD and field theory (Figs.~\ref{fig:fig4}b,c). More generally, numerical solutions of our field theory show instances of all dynamical processes seen in simulations. Thus, the kymograph in Fig.~\ref{fig:fig4}d shows examples of Ostwald ripening and fusion, whereas in steady state we find hopping and fission if $\sigma_{\eta}$ is sufficiently large (Fig.~\ref{fig:S9}).

\section{Conclusions} 

In summary, we have studied the dynamics of polymer braids under tension by numerical simulations. By varying the applied force and the catenation (linking) number between the two chains, we mapped out the transition between the straight braid and plectonemic phase. Our main result is the characterisation of dynamical events and kinetic regimes deep in the plectonemic phase. We find that plectoneme formation shares some features of phase separation in liquid-gas systems and proceeds via a combination of Ostwald-like ripening and fusion events. At late times, there is a balance of merging and breakup of plectonemes which leads to arrest of coarsening with selection of a typical plectoneme size. We also provide evidence of plectonemic hopping, where writhed regions seemingly unravel at one location to appear somewhere else along the braid. 
  
We showed that our numerical data can be explained by a stochastic field theory which models twist and writhe interconversion in our polymer braid. This theory suggests that the growth of plectonemes can be described by a power law, where the exponent is $\sim 1/3$, as befits diffusive phase separation in a liquid-gas system. It also reproduces hopping and concurs with our BD simulations in finding that sufficiently strong noise (i.e., proximity to the transition) is crucial to observe this dynamical mechanism. Both BD and field theory suggest that this phenomenon is due to stochastic nucleation of an additional plectoneme and subsequent competition between the nascent plectoneme and another one in the chain. As the latter unravels, the result is an apparent hopping. 
In the future, it would be of interest to study quantitatively via single molecule techniques the dynamics and growth laws of plectonemes in DNA and polymer braids, to confirm whether the diffusive liquid-gas phase separation outlined here  is confirmed experimentally. 

\section{Acknowledgments}
We thank ERC (CoG 648050 THREEDCELLPHYSICS) for funding.

\section*{Appendix}
In this Appendix we discuss the details of the continuum reaction-diffusion model whose results have been described in the text. The model consists of two coupled dynamical equations for twist and writhe, which we write down here again for convenience,
\begin{eqnarray}\label{meanfieldtheory}
\frac{\partial {\rm Tw}}{\partial t} & = & D_t \frac{\partial^2 {\rm Tw}}{\partial x^2} - g({\rm Tw},{\rm Wr}) \\ \nonumber
\frac{\partial {\rm Wr}}{\partial t} & = & D_w \frac{\partial^2 {\rm Wr}}{\partial x^2} - \kappa \frac{\partial^4 {\rm Wr}}{\partial x^4}
+g({\rm Tw},{\rm Wr}) \\ \nonumber
g({\rm Tw},{\rm Wr}) & = &  \alpha {\rm Tw}   - \beta {\rm Wr} + \gamma \Theta\left(|{\rm Tw} + {\rm Wr}| - {\sigma}_0\right) {\rm Wr} + \sigma_{\eta} \eta(x,t).
\end{eqnarray}
In Eq.~\ref{meanfieldtheory}, $D_t$ and $D_w$ are the effective diffusion coefficients of twist and writhe respectively, $\theta(x)$ is the Heaviside function, while $\kappa>0$ introduces a surface tension in writhed domains. The reaction terms, called $g$ and $-g$ in the equations for twist and writhe, are equal and opposite so that the linking number, or local catenation number $\sigma(x,t)$ -- equal to ${\rm Tw}(x,t)+{\rm Wr}(x,t)$, is globally conserved at all times. 
We considered $\alpha \ge \beta\ge \gamma\ge 0$, while $\sigma_0$ is the local linking/catenation number at which plectonemes appear. Finally, $\sigma_{\eta}$ controls the strength of noise, and $\eta(x,t)$ is a white noise with zero mean and variance equal to 
\begin{equation}
\langle \eta(x,t) \eta(x',t')\rangle=\delta(x-x') \delta(t-t'),
\end{equation}
where $\langle \cdot \rangle$ denotes averaging over different noise realisations. In the following calculations we set $\kappa=0$ for simplicity -- this does not affect our analytical treatment. 

\begin{figure}[!]
\centering \includegraphics[width=0.4\textwidth]{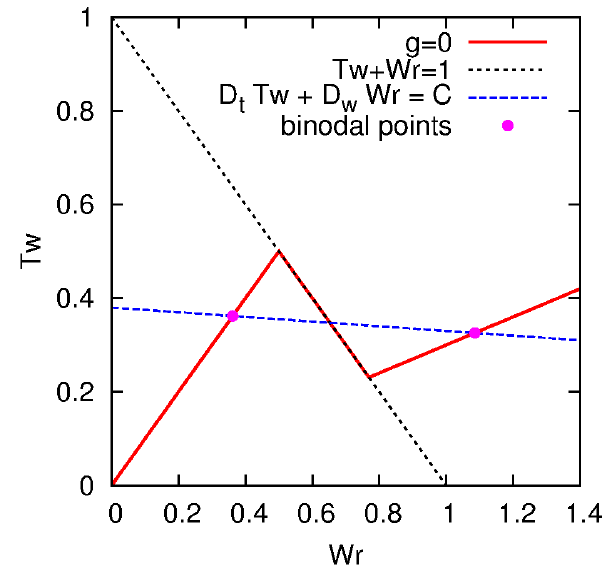}
\caption{
\textbf{Phase separation in a reaction-diffusion model for plectoneme formation.} 
Plot showing the curve corresponding to $g=0$ (red line), together with the straight line $D_t {\rm Tw}+D_w{\rm Wr}=C$ (blue line). The intersection between the two identifies possible stationary points: of which the two magenta filled circles correspond to stable solutions (these are binodal points in phase separation theory). The dotted line correspond to the line ${\rm Tw}+{\rm Wr}=1$. The existence of two stable solutions signifies that the system phase separate into a highly writhed (``plectonemic'') and a weakly writhed (``straight'') phase to keep the total linking number conserved. Parameters for this diagram are: $\alpha=\beta=0.1$, $\gamma=0.07$, ${\sigma}_0=1$, $C/D_t=0.38$, $D_w/D_t=0.05$.}
\label{fig:S8}
\end{figure}

Let us now assume that Eq.~\ref{meanfieldtheory} admits a stationary solution, where twist and writhe are constant in steady state. This solution will be such that the reaction term vanishes, $g=0$. Additionally, by summing the two equations we obtain
\begin{equation}
\frac{\partial \left({\rm Tw}+{\rm Wr}\right)}{\partial t} = \frac{\partial}{\partial x}\left[D_t \frac{\partial {\rm Tw}}{\partial x}+D_w\frac{\partial {\rm Wr}}{\partial x}\right].
\end{equation}   
Consequently, for a stationary solution, and if there is no twist or writhe flux at the boundaries (i.e., $\frac{\partial {\rm Tw}}{\partial x}=\frac{\partial {\rm Wr}}{\partial x}=0$ there), we obtain that $D_t {\rm Tw}+D_w{\rm Wr}=C$, where $C$ is a constant which is determined by the initial condition at $t=0$. Intersecting the curve $g=0$ with the straight line $D_t {\rm Tw}+D_w{\rm Wr}=C$ gives the possible stationary solutions of the system. Notably, for a range of values of $C$ the straight line  $D_t {\rm Tw}+D_w{\rm Wr}=C$ intersects $g=0$ in three points, and the two filled circles in Fig.~S\ref{fig:S8} denotes the two stable points -- these are the analog of binodal points for a phase separating system. [Note that we work in the quadrant with ${\rm Tw}>0$ and ${\rm Wr}>0$ without loss of generality, as the rest of the behaviour can be obtained using symmetry.]

The presence of two distinct stable stationary points means that the system will phase separate into two (or more) domains, with a different stable point in each pair of neighbouring domains, and where domain size is set by the global constraint that the total linking number is globally conserved. This argument is the same as that used in~\cite{Frey_Nat_Phys_2018} to demonstrate the formation of spatially varying stationary patterns in MinD protein systems in bacteria. It is important to note that in order for two stable solutions to exist -- equivalently, in order for phase separation to arise -- we need $D_t>D_w$, which is the case for DNA.

\begin{figure*}[!]
\centering \includegraphics[width=0.95\textwidth]{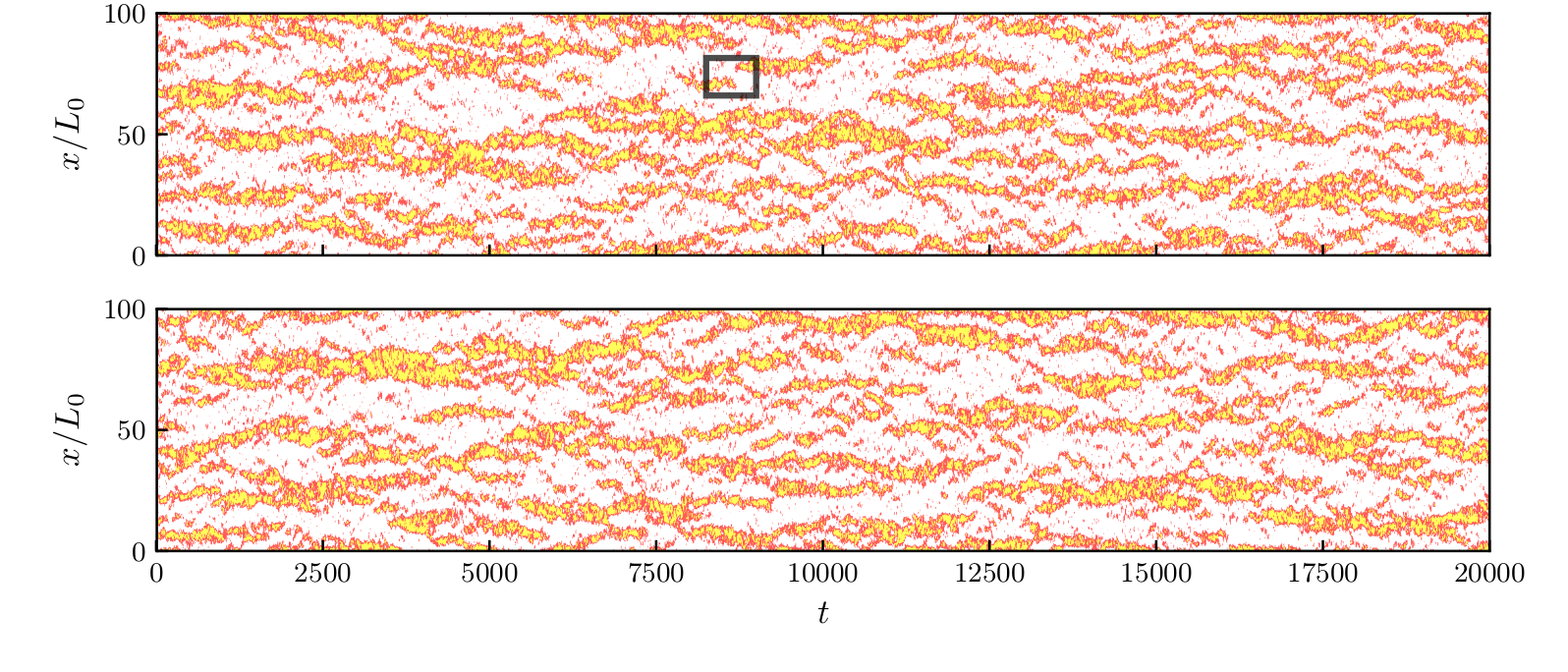}
\caption{
\textbf{Kymographs for the reaction-diffusion model.}
(a,b) Kymographs showing two realisation of the dynamics of plectonemes in the continuum model when noise strength is larger than in the main text. A hopping event in (a) is highlighted by a box. The values of the  parameters used for these realisations are: $D_t=2$, $D_w=0.01$ $\alpha=\beta=0.1$, $\gamma=0.07$, ${\sigma}_0=1$, $\kappa=0.01$, $\sigma_{\eta}=0.104$, $L_0=100$.}
\label{fig:S9}
\end{figure*}

It is also instructive to write down an effective equation for the local linking/catenation number, $\sigma={\rm Tw}+{\rm Wr}$, which corresponds to a globally conserved variable. 
To this end, we can add and subtract the equations for twist and writhe in Eq.~\ref{meanfieldtheory}, to obtain
\begin{eqnarray}\label{meanfieldtheory2}
\frac{\partial \sigma}{\partial t} & = & \frac{D_t+D_w}{2}\frac{\partial^2 \sigma}{\partial x^2} +\frac{D_t-D_w}{2} \frac{\partial^2 \delta}{\partial x^2} \\ 
\nonumber
\frac{\partial \delta}{\partial t} & = & \frac{D_t-D_w}{2}\frac{\partial^2 \sigma}{\partial x^2} +\frac{D_t+D_w}{2} \frac{\partial^2 \delta}{\partial x^2} + 2g,
\end{eqnarray}
where we called $\delta={\rm Tw}-{\rm Wr}$.

Now, we perform a gradient expansion in the second equation to express $\delta$ as a function of $\sigma$. To zero-th order in the gradients, the relation between $\delta$ and $\sigma$ is obtained by setting $g=0$, namely, 
\begin{equation}
\delta = \frac{\beta-\alpha-\gamma \Theta\left(|\sigma| - \sigma_0\right)}{\beta+\alpha -\gamma\Theta\left(|\sigma| - \sigma_0\right)} \sigma.
\end{equation}
Plugging this into the equation for $\sigma$, we find that it can be written as an effective ``model B'' dynamics (in the terminology of~\cite{chaikinlubensky}), which is the relevant equation of motion for a globally conserved order parameter. The equation explicitly reads as follows,
\begin{equation}
\frac{\partial \sigma}{\partial t}= M \frac{\partial^2 \mu_{\rm eff}}{\partial x^2}\equiv M \frac{\partial^2}{\partial x^2} \left(\frac{\partial f_{\rm eff}}{\partial \sigma}\right),
\end{equation}
with $M$ a mobility (which we assume constant), $\mu_{\rm eff}$ an effective chemical potential, and $f_{\rm eff}$ an effective free energy density. The latter is given by:
\begin{eqnarray}\label{feff} 
 M f_{\rm eff} & = & A+\left[\frac{D_t+D_w}{2}-\frac{D_t-D_w}{2}\frac{\alpha-\beta}{\alpha+\beta}\right]\sigma^2\nonumber \\
 &\equiv& A+C_1\sigma^2 \qquad {\rm if} \qquad \sigma<{\sigma}_0 \\ \nonumber
M f_{\rm eff} & = & \left[\frac{D_t+D_w}{2}-\frac{D_t-D_w}{2}\frac{\alpha-\beta+\gamma}{\alpha+\beta+\gamma}\right] \sigma^2 \nonumber \\
&\equiv& C_2\sigma^2 \qquad \qquad \, {\rm if} \qquad \sigma>{\sigma}_0. 
\end{eqnarray}
In Eq.~\ref{feff}, $A$ is a constant to ensure continuity of $f_{\rm eff}$ at $\sigma={\sigma}_0$, whereas given the choices of $\alpha$, $\beta$, $\gamma$ $C_1>C_2>0$. Therefore the effective free energy is piecewise quadratic, with a singularity at ${\sigma}_0$. This function is non-convex so that a common tangent construction predicts phase separation, in line with our previous analysis of the reaction-diffusion model. We note that an effective free energy with the same functional form was proposed by Marko in~\cite{Mark_PRE_2007} to study the thermodynamics of plectoneme formation in supercoiled DNA under tension. 
Whilst our system of equations leads to phase separation and whilst we can compute the value of the twist and writhe of the coexisting states, a linear stability analysis as in~\cite{Frey_Nat_Phys_2018} shows that the only point where the system undergoes spinodal decomposition is $\sigma={\rm Tw}+{\rm Wr}={\sigma}_0$, which can also be seen in our effective free energy, as $\frac{\partial^2 f_{\rm eff}}{\partial x^2}$ is always positive except at $\sigma={\sigma}_0$ -- this is due to the choice of a singular reaction term (or equivalently effective free energy).
As shown in Fig.~\ref{fig:fig4}, the dynamics of the continuum model is similar to that of our molecular dynamics simulations. Additional kymographs are shown in Fig.~\ref{fig:S9} for larger noise strength. These show some examples of hopping. The mechanism leading to these is similar to that identified in the molecular dynamics simulations, with a plectoneme nucleating approximately at the same time when one of similar size disappears -- the combined effect is an apparent hopping of the plectoneme along the chain.

\bibliographystyle{apsrev4-1}
\bibliography{biblio_braid}

\begin{thebibliography}{23}
\bibitem [{Not()}]{Nota1}%
  \BibitemOpen
  \href@noop {} {}\bibinfo {note} {In practice, a good choice of the thresholds
  for our simulations was $R_{\mbox{\scriptsize{max}}}=3 \, \sigma$ and
  $\Delta_p \simeq \frac{4 \pi \sigma}{\sigma} \simeq 13$ beads (as the
  smallest possible plectoneme is represented by a perfect circle with diameter
  $2 \sigma$, the thickness of the braid)}\BibitemShut {NoStop}%
\end{thebibliography}


\end{document}